\documentclass[notitlepage,nofootinbib,preprintnumbers,amssymb,superscriptaddress,eqsecnum]{revtex4-1}
\usepackage{amsfonts,amssymb,amsmath,graphicx,color,bm}
\definecolor{ultramarine}{rgb}{0.07, 0.04, 0.56}
\definecolor{cadmiumgreen}{rgb}{0.0, 0.42, 0.24}
\definecolor{indigo(dye)}{rgb}{0.0, 0.25, 0.42}
\usepackage[linktocpage=true]{hyperref}
\hypersetup{
colorlinks=true,
citecolor=ultramarine,
linkcolor=cadmiumgreen,
urlcolor=indigo(dye),
}

\newcommand{\f}[2]{\frac{#1}{#2}}  
\newcommand{\mk}[1]{\left( #1 \right)}  
\newcommand{\kk}[1]{\left[ #1 \right]}  
\newcommand{\ck}[1]{\left\{ #1 \right\}}  
\newcommand{\be}{\begin{equation}}  
\newcommand{\ee}{\end{equation}}
\newcommand{\Mpl}{M_{\rm Pl}}
\newcommand{\e}{\epsilon}
\renewcommand{\d}{\delta}
\newcommand{\pa}{\partial}

\begin{document}

\title{  
Constant-roll inflation in scalar-tensor gravity
}

\author{Hayato Motohashi}
\affiliation{Center for Gravitational Physics, Yukawa Institute for Theoretical Physics, Kyoto University, Kyoto 606-8502, Japan}

\author{Alexei A.\ Starobinsky}
\affiliation{L. D. Landau Institute for Theoretical Physics RAS, Moscow 119334, Russia}
\affiliation{Research Center for the Early Universe (RESCEU),
Graduate School of Science, The University of Tokyo, Tokyo 113-0033, Japan}

\preprint{YITP-19-81}

\begin{abstract}
We generalize the notion of constant-roll inflation earlier introduced in General Relativity (GR) and $f(R)$ gravity to inflationary models in more general scalar-tensor gravity. A number of novel exact analytic solutions for a FLRW spatially flat cosmological background is found for this case. All forms of the scalar field potential and its coupling to gravity producing the exact de Sitter solution, while the scalar field is varying, are presented. In the particular cases of induced gravity and GR with a non-minimally coupled scalar field, all constant-roll inflationary solutions are found. In the former case they represent power-law inflation, while in the latter case the solution is novel and more complicated. Comparison of scalar perturbations generated during such inflation in induced gravity with observational data shows that the constant-roll parameter should be small, similar to constant-roll inflation in GR and $f(R)$ gravity. Then the solution reduces to the standard slow-roll one with small corrections.  
\end{abstract}

\maketitle

\section{Introduction}

Slow-roll inflationary models predict observables consistent with the latest observations.
Yet, it is possible to consider more general models observationally viable by replacing the assumption of the inflaton slow roll to constant roll.
As a generalization of the standard slow-roll condition, the constant-roll condition $\ddot\phi=\beta H\dot\phi$ was first employed in~\cite{Martin:2012pe}, where $\phi$ is the inflaton, and $H\equiv \dot a/a$ is the Hubble parameter.
The constant-roll condition with small $|\beta|$ is compatible with the slow roll of the inflaton, whereas a special value $\beta=-3$ amounts to the ultra-slow-roll inflation~\cite{Tsamis:2003px,Kinney:2005vj}. 
Later, using the Hamiltonian-Jacobi approach, the analytic solution of the inflationary dynamics and the potential was found in \cite{Motohashi:2014ppa} for canonical inflation, which was explicitly shown to be compatible with the observational constraint for $\beta\approx 0.015$~\cite{Motohashi:2014ppa,Motohashi:2017aob} (see also \cite{GalvezGhersi:2018haa}).
This solution represents an inflationary model having two free parameters ($\beta$ and the initial inflaton position $\phi_{\rm ini}$) with an assumption of a smooth transition to reheating phase.
The constant-roll inflation with a different value of $\beta$ has been also extensively studied recently in~\cite{Morse:2018kda,Gao:2019sbz,Lin:2019fcz,Motohashi:2019rhu}.

Similar analytic solution exactly satisfying the constant-roll condition in the Jordan frame $\ddot f'=\beta Hf'$ was also obtained in $f(R)$ gravity in \cite{Motohashi:2017vdc}, in which the master first-order differential equation for $H$ in terms of the curvature $R$ in the Jordan frame was also derived for general $f(R)$ model.
While the function $f(R)$ for exact constant-roll inflation is obtained in the parametric form, it can be shown that focusing on small $|\beta|$, it is close to $R^2$ according to the general theorem valid for any slow-roll inflation model in $f(R)$ gravity with arbitrary, but small values of the scalar power spectrum slope $n_s(k)-1$ and the tensor-to-scalar ratio $r(k)$~\cite{Appleby:2009uf}. For the value $n_s(k)\approx 0.965$ at cosmological scales preferred by observational data, the relevant slow-roll models satisfying this condition are the $R+R^2$ inflationary model~\cite{Starobinsky:1980te}, or the $R+R^p$ model with $p\approx 2$~\cite{Motohashi:2014tra}.
However, it remains unclear if constant-roll inflation exists in more general framework with non-minimal coupling between the inflaton and gravity.

So, in this paper we generalize the notion of constant-roll inflation to more general scalar-tensor theory of gravity.
The rest of the paper is organized as follows.
In \S\ref{sec:cr}, we impose the constant-roll condition and provide basic equations for dynamics of background space-time and perturbations.
In \S\ref{sec:ds}, we find the constant-roll model in scalar-tensor gravity allows an exact de Sitter solution as an analytic solution of equations of motion.
In \S\ref{sec:ig}, we focus on the induced gravity and investigate the constant-roll inflationary scenario in detail to show that there exists a viable parameter region that can accommodate the latest observational constraints.
In \S\ref{sec:st} and \S\ref{sec:exp}, we provide analytic solutions for the particular cases of GR driven by a non-minimally coupled scalar field with a potential.
\S\ref{sec:con} is devoted to the conclusion.

\section{Constant-roll dynamics}
\label{sec:cr}

\subsection{Background}

Let us consider the following scalar-tensor theory with a non-minimal coupling of a scalar field to gravity: 
\be \label{action} S=\int d^4x \sqrt{-g} \kk{\f{1}{2}f(\phi)R -\f{\sigma}{2} \pa_\mu\phi\pa^\mu\phi - V(\phi)} , \ee
where $\sigma=\pm 1$ and we use the natural unit where $\Mpl\equiv (8\pi G)^{-1/2}=1$.
This action is equivalent to the Brans-Dicke theory 
\be S=\int d^4x \sqrt{-g} \kk{\f{1}{2}\Phi R -\f{\omega_{\rm BD}(\Phi)}{2\Phi} \pa_\mu\Phi\pa^\mu\Phi - V(\phi(\Phi))} , \ee
via a redefinition of the scalar field
\be \Phi=f(\phi), \quad \omega_{\rm BD}(\Phi) = \f{\sigma f}{f'^2} . \ee
A requirement that the scalar field mediates an attractive force in the weak field limit yields the condition $(6+4\omega_{\rm BD})^{-1}>0$, namely~\cite{Fujii:2003pa} 
\be \label{nogh} -\f{3}{2} < \omega_{\rm BD} < \infty. \ee
Therefore, in general both signs $\sigma=\pm 1$ can be free from ghost so long as $\omega_{\rm BD}$, or $f(\phi)$ satisfies the condition~\eqref{nogh}.

With the spatially flat Friedmann-Lema\^itre-Robertson-Walker (FLRW) background metric, $ds^2=-dt^2+a(t)^2\delta_{ij}dx^idx^j$,
the equations of motion of the action~\eqref{action} are given by
\begin{align} \label{eom}
&3 f H^2 = \f{\sigma}{2} \dot\phi^2 + V - 3 H f' \dot\phi ,\notag\\
&-2 f \dot H = (\sigma + f'') \dot\phi^2 + f' (\ddot\phi - H \dot\phi), \notag\\
&\sigma(\ddot\phi + 3 H \dot\phi) + V' - 3 f' (2 H^2 + \dot H) = 0 ,
\end{align}
where $H\equiv \dot a/a$ is the Hubble parameter, and a dot and a prime denote a derivative with respect to $t$ and $\phi$, respectively. 
They recover the standard equations for the canonical inflation for $f=1$ and $\sigma=+1$.

We are interested in exact solutions that satisfy the constant-roll condition
\be \label{conroll} \ddot\phi = \beta H \dot\phi, \ee
where $\beta$ is a constant.\footnote{Note that the condition~\eqref{conroll} is not conformally invariant. Therefore, a constant-roll inflationary solution in the original Jordan frame does not transform into a constant-roll inflation in GR in the dual Einstein frame, as was shown in \cite{Motohashi:2017vdc}.  }
Following~\cite{Motohashi:2014ppa}, we adopt the Hamiltonian-Jacobi approach and regard $H=H(\phi)$ to derive constant-roll exact solutions. 
Plugging the constant-roll condition~\eqref{conroll} and $\dot H = \dot\phi H'(\phi)$ to the second equation of \eqref{eom}, we obtain
\be \label{dphi0} \dot\phi[(\sigma + f'') \dot\phi + (\beta-1) f'H + 2f H' ]=0 . \ee

First, let us consider the case where $\dot\phi\neq 0$ and $\sigma+f''\neq 0$ are satisfied.
We can then express $\dot\phi$ as a function of $\phi$ as
\be \label{dphi} \dot\phi = \f{1}{\sigma+f''} [ (1-\beta) f'H - 2fH' ] . \ee
By taking a time derivative of \eqref{dphi} and plugging the constant-roll condition~\eqref{conroll} once more, we obtain a differential equation for $H(\phi)$ as
\be \label{dH} \beta H = \f{\pa}{\pa\phi} \kk{ \f{1}{\sigma+f''} [ (1-\beta) f'H - 2fH' ] } . \ee
This equation is a natural generalization of the differential equation $\beta H = - 2 H''$ obtained in~\cite{Motohashi:2014ppa} for the canonical constant-roll inflation, which can be recovered by setting $f=1$ and $\sigma=+1$.

Depending on the functional form of $f(\phi)$, we can obtain several branches of the constant-roll exact solutions of the differential equation~\eqref{dH} of $H(\phi)$.
We shall discuss each branch in \S\ref{sec:ds}--\ref{sec:st}.
For the following, we consider general solution.

In general, for a given function $f(\phi)$ the differential equation \eqref{dH} yields two independent solutions for $H(\phi)$.
Let them be $h_1(\phi)$ and $h_2(\phi)$. 
The Hubble parameter is then formally given by a linear combination
\be \label{Hsol} H(\phi) = M[c_1h_1(\phi)+c_2h_2(\phi)] , \ee  
where $c_1,c_2$ are dimensionless constants of integration
and $M$ is the energy scale of inflation, which is fixed by the CMB normalization of the amplitude of the scalar power spectrum.
Therefore, in the most of the present paper, we set $M=1$ for simplicity.
The time evolution of the scalar field in terms of $t$ can be obtained by solving \eqref{dphi} with \eqref{Hsol}.

Once $H$ and then $\dot\phi$ are given as a function of $\phi$ by \eqref{Hsol} and \eqref{dphi}, 
we can derive the potential $V=V(\phi)$.
From the first equation of \eqref{eom} we obtain
\be \label{pot} V(\phi) = 3 f H^2 - \f{\sigma}{2} \dot\phi^2 + 3 H f' \dot\phi , \ee
Plugging \eqref{Hsol} with $M=1$ to \eqref{dphi} we obtain $\dot\phi$ as a function of $\phi$, 
and then \eqref{pot} reads
\be \label{Vsol} V(\phi) = c_1^2 v_1(\phi) + c_2^2 v_2(\phi) + 2 c_1 c_2 v_3(\phi), \ee
where
\begin{align} \label{veqs}
v_1(\phi) &= g_1 h_1^2 + 2 g_2 h_1h'_1 + g_3 h_1'^2, \notag\\
v_2(\phi) &= g_1 h_2^2 + 2 g_2 h_2h'_2 + g_3 h_2'^2, \notag\\
v_3(\phi) &= g_1 h_1h_2 + g_2 (h_1h'_2 + h'_1h_2) + g_3 h'_1h'_2,
\end{align}
and
\begin{align} \label{geqs}
g_1(\phi) &= 3f  +\f{(1 - \beta) f'^2 [(5 + \beta) \sigma + 6 f'']}{2 (\sigma + f'')^2}, \notag\\
g_2(\phi) &= -\f{f f' [(2 + \beta) \sigma + 3 f'']}{(\sigma + f'')^2}, \notag\\
g_3(\phi) &= -\f{2 \sigma f^2}{(\sigma + f'')^2} .
\end{align}

As mentioned above, \eqref{dphi} is based on the assumptions $\dot\phi\neq 0$ and $\sigma+f''\neq 0$.
On the other hand, we can consider a special case where $\sigma+f''= 0$ is satisfied.
In this case, we have
\be \label{sol2} f(\phi)=  -\f{\sigma \phi^2}{2} + f_1 \phi + f_2 , \quad
H(\phi) = M f(\phi)^{(1-\beta)/2}, \ee
where $f_1,f_2$ are integration constants, and 
the Hubble parameter is obtained by solving $(\beta-1) f'H + 2f H' =0$ from \eqref{dphi0}.
To obtain the potential, one needs to solve \eqref{conroll} for $\dot\phi$ and plug it into \eqref{pot}.
While in general this process cannot be performed analytically, 
it is possible for a special case, as we shall see in \S\ref{sec:ds}.

\subsection{Perturbation}

Before proceeding details of each solution, let us summarize the tensor and scalar perturbations of the action~\eqref{action}.
We consider metric perturbation $h_{\mu\nu} = g_{\mu\nu} - \bar g_{\mu\nu}$ around the flat FLRW metric $\bar g_{\mu\nu}$ in the unitary gauge where perturbation of the scalar field $\delta \phi = 0$, which coincides with the comoving gauge.
After integrating out nondynamical scalar perturbations,
for the tensor and scalar perturbations defined by
\be h_{ij} = a^2 (e^{2\zeta}\delta_{ij} + \gamma_{ij} ),  \ee
the quadratic actions are given by
\begin{align} 
S_{2,t} &= \int d^4x~ \sum_{\lambda=+,\times} \f{a^3 b_t}{4c_t^2} \mk{ \dot\gamma_\lambda^2 - \f{c_t^2k^2}{a^2} \gamma_\lambda^2} , \notag\\
S_{2,s} &= \int d^4x~ \f{a^3 b_s \e_H}{c_s^2} \mk{ \dot\zeta^2 - \f{c_s^2k^2}{a^2}\zeta^2}, 
\end{align}
where $\epsilon_H\equiv -\dot H/ H^2$ and 
\begin{align} \label{bc}
b_t &= f , \hspace{16mm} 
c_t^2 = 1 ,\notag\\
b_s &= f + \f{f' \dot\phi}{H \e_H} ,\quad
c_s^2 = \f{2 H (f H \e_H + f' \dot\phi)}{\sigma \dot\phi^2} . 
\end{align}
This notation allows us to capture the difference from the canonical case explicitly by the sound speed $c_i$ and normalization factor $b_i$ and to highlight their role in slow roll expansion for a wide range of theories~\cite{Motohashi:2015hpa,Motohashi:2017gqb,Motohashi:2020wxj}.

Solving the Euler-Lagrange equation at the superhorizon limit, we obtain the superhorizon solutions as 
\begin{align} \label{shsol}
\gamma_{+,\times} &\approx C_t + D_t \int dt \f{c_t^2}{a^3 b_t} ,\notag\\
\zeta &\approx C_s + D_s \int dt \f{c_s^2}{a^3 b_s \e_H} ,
\end{align}
where $C_i, D_i$ are integration constants.
While in the standard inflationary scenario the second modes are decaying mode, it is actually not always the case.
For instance, in the ultra-slow-roll inflation, the curvature perturbation is growing on superhorizon scales~\cite{Kinney:2005vj}, which leads to the violation of the non-Gaussianity consistency relation~\cite{Namjoo:2012aa,Martin:2012pe}. 
For the following, we shall focus on the case where the second modes are decaying modes.

\section{Exact de Sitter solution}
\label{sec:ds}

In this section, we derive an exact de Sitter constant-roll solution in scalar-tensor theory.
The exact de Sitter solution does not exist in the canonical constant-roll model.
From the point of view of the application to the early Universe, 
since the primordial perturbations are not generated during exact de Sitter regime, 
it is necessary to introduce other mechanism to produce the primordial perturbation such as curvaton.

Plugging $H=M=\text{const.}$ to \eqref{dH} yields
\be \beta=(1-\beta)\f{\pa}{\pa\phi}\mk{\f{f'}{\sigma+f''}}. \ee
A special case $\beta=1$ corresponds to $\sigma+f''=0$, for which \eqref{dH} does not apply as mentioned above.
Hence, below we shall focus on the case $\beta\ne 1$ first.
We shall address de Sitter solution for $\beta=1$ at the end of this section, 
and show that while the derivation is slightly different, de Sitter solution for $\beta= 1$ coincides with the solution derived for $\beta\ne 1$ with substitution $\beta= 1$.

For $\beta\ne 1$ we obtain
\be \label{feq} \f{f'}{ \sigma + f''} = \f{\beta}{1-\beta}\phi, \ee
where we absorbed a constant of integration into a redefinition of the scalar field.
From \eqref{dphi} we obtain $\dot\phi=\beta M\phi$ and hence
\be \phi \propto e^{\beta M t} .\ee
Therefore, the assumption $\dot \phi\neq 0$ is satisfied. 
Note that this branch shows up only when $f'\neq 0$, i.e.\ when the scalar field non-minimally couples to gravity.
The de Sitter branch with $\dot\phi\neq 0$ is not possible in the case of GR with a canonical scalar field.
For a given scalar-tensor theory with $f(\phi)$ satisfying the differential equation~\eqref{feq}, the de Sitter solution with time-evolving scalar field $\phi \propto e^{\beta M t}$ exists.

The simplest example of such case was known long ago~\cite{Barrow:1990nv,Sahni:1998at}: 
a massive non-minimally coupled scalar field with 
\be \label{BS} \sigma=+1, \quad f(\phi)=1-\xi\phi^2, \quad V(\phi)=\f{m^2}{2}\phi^2. \ee
Indeed, \eqref{feq} is satisfied with
\be \label{betasol} \beta=\f{2\xi}{4\xi-1}, \ee
and then the potential \eqref{pot} is given by
\be \label{potsol} V(\phi) = -\f{\xi(6\xi-1)(16\xi-3)}{(4\xi-1)^2}M^2\phi^2 + 3M^2 . \ee
Therefore, the model with $V(\phi)=m^2\phi^2/2$ corresponds to 
\be \f{M^2}{m^2}=-\f{(4\xi-1)^2}{2\xi(6\xi-1)(16\xi-3)} , \ee
neglecting the constant contribution $3M^2$.
The mass squared of the potential \eqref{potsol} is nonnegative for $\xi\leq 0$ or $1/6\leq \xi \leq 3/16$.
For recent interest to this topic, see e.g.\ \cite{Sami:2012uh}.
Also, note that for the special case of the model~\eqref{BS} with $\beta=-1$, we obtain the conformal coupling $\xi=1/6$ and constant potential $V(\phi)=3M^2$, which was studied in \cite{Kofman:2007tr}.

Let us derive general solution of the differential equation~\eqref{feq}.
Using the derivation of general constant-roll solution in \S\ref{sec:cr}, we can find all solutions with $H=M=$~const.\ and $\phi \propto e^{\beta M t}$ for the coupling $f(\phi)$ and potential $V(\phi)$.
The coupling $f(\phi)$ is determined by solving the differential equation~\eqref{feq} as 
\be \label{dsf} f(\phi) = \sigma \gamma \phi^2 + d_1 \beta \phi^{1/\beta} + d_2 , \ee
with
\be \gamma \equiv \f{\beta}{2 (1 - 2 \beta)}, \ee
and $d_1, d_2$ are constants of integration.  
Note that $d_1, d_2$ are arbitrary constants including zero whereas we remind $\sigma=\pm 1$.
The model~\eqref{BS} corresponds to a special case with $d_1=0$, $d_2=1$, $\sigma=+1$, $\gamma=-\xi$, which is consistent with \eqref{betasol}.
Plugging \eqref{dsf} and $\dot\phi=\beta M\phi$ into \eqref{pot}, we obtain the potential as  
\begin{align} 
\label{dspot}
V(\phi) 
&= M^2 \mk{ \f{\sigma \gamma  (6 \gamma + 1 ) (16 \gamma + 3 )}{(4 \gamma + 1)^2} \phi^2 + 6 d_1 \beta \phi^{1/\beta} + 3 d_2  } . 
\end{align}
It is also clear that \eqref{dspot} coincides with \eqref{potsol} for the case of the above parameter set.

Finally, let us consider the exceptional case $\beta=1$.
In this case, instead of \eqref{feq}, we can use \eqref{sol2}, and obtain 
\begin{align} \label{dSb1}
f(\phi) &=  -\f{\sigma \phi^2}{2} + f_1 \phi + f_2 , \notag\\
V(\phi) &= M^2 (-5 \sigma \phi^2 + 6 f_1 \phi + 3 f_2 ).
\end{align}	
which actually coincide with \eqref{dsf} and \eqref{dspot} with $\beta=1$.
Therefore, the de Sitter solution \eqref{dsf} and \eqref{dspot} also apply to the case $\beta=1$.

\section{Induced gravity}
\label{sec:ig}

We proceed to detailed analysis of specific theories, derive constant-roll exact solutions by solving the differential equation~\eqref{dH}, and study its observational predictions.
In this section we focus on 
the induced gravity with
\be \label{find} f(\phi)=\xi\phi^2, \ee
where $\xi$ is a dimensionless constant.
Since $\omega_{\rm BD}(\Phi) = \sigma f/f'^2= \sigma /4\xi$ in this case, we focus on the ghost-free parameter region \eqref{nogh}, i.e.\
\be \label{noghost-ind} -6 < x < \infty, \ee
where we defined $x\equiv \sigma/\xi$.
The assumption $\sigma+f''\ne 0$ reads $\sigma+2\xi\ne 0$ or $x\ne -2$.

With \eqref{find}, the differential equation~\eqref{dH} reads
\be \label{dHind} H'' + (1+\beta)\f{H'}{\phi} + \kk{ -1 + \beta \mk{2 + \f{x}{2}} }\f{H}{\phi^2} = 0 .  \ee
The two independent solutions of \eqref{dHind} are $\phi^{(-\beta\pm p)/2}$ with
\be \label{defp} p\equiv 
\sqrt{(\beta-2)^2 - 2 \beta ( 2 + x) }. \ee
Below we consider two cases: $(\beta-2)^2 - 2 \beta ( 2 + x) \gtrless 0$ and denote them Case~1 and Case~2, respectively.
Clearly, $p$ is real for Case 1, whereas $p$ is pure imaginary for Case 2. 
The conditions simplify as
\begin{align}
\label{realp} \text{Case 1}: & \quad  \beta \kk{ x - \mk{-4 + \f{2}{\beta} + \f{\beta}{2} } } < 0, \\
\text{Case 2}: & \quad  \beta \kk{ x - \mk{-4 + \f{2}{\beta} + \f{\beta}{2} } } > 0.	
\end{align}

The potential is given by \eqref{Vsol} with \eqref{veqs} and \eqref{geqs}, the latter of which plugging \eqref{find} reads
\begin{align} \label{gind}
g_1(\phi)&= \xi \kk{ 3 + \f{2 (1 - \beta)  [12  + (5 + \beta) x]}{(2  + x)^2} }\phi^2,\notag\\
g_2(\phi)&= -\f{2 \xi [6  + (2 + \beta) x] }{(2  + x)^2} \phi^3 ,\notag\\
g_3(\phi)&= -\f{2 \sigma}{(2 + x)^2} \phi^4 .	
\end{align}
For simplicity, we focus on the scenario where inflation occurs at the positive region $\phi>0$.

\subsection{Case 1, solution 1}

Let us begin with the Case 1 defined by \eqref{realp}.
In this case, the analytic solution for the Hubble parameter $H(\phi)$ is given by \eqref{Hsol}, i.e.\ a linear combination of 
\be \label{sol-realp} h_1(\phi) = \phi^{(-\beta - p)/2}, \quad 
h_2(\phi)=\phi^{(-\beta + p)/2}, \ee
with real integration constants $c_1,c_2$.
The relation between $\phi$ and $t$ can be obtained by solving \eqref{dphi}, 
which is expressed in terms of the Gauss' hypergeometric function ${}_2F_1$.
Finally the potential is given by \eqref{Vsol} with \eqref{gind} and \eqref{sol-realp}.

For a particular solution, the potential takes a simple form.
Below we focus on a particular solution
\be \label{h1} H(\phi) = h_1 (\phi) = \phi^{(-\beta - p)/2} . \ee
For another particular solution, 
\be \label{h2} H(\phi) = h_2 (\phi) = \phi^{(-\beta + p)/2} , \ee
we shall obtain similar expressions with a replacement $p \to -p$.

For the particular solution \eqref{h1} 
we obtain
\begin{align} \label{sol1-ind}
V(\phi) &= \f{\xi[12 (3 - \beta + p) 
+ \{ 20 - \beta^2 + p (4 + \beta) \} x
+ (3 + \beta)x^2]}{(2 + x)^2 }  \phi^{2 - \beta - p}  , \notag\\
\phi(t) &= \mk{ \f{(\beta + p) (2 - \beta + p) }{2 (2 + x )} t }^{2/(\beta + p)} ,\notag\\
H(t) &= \mk{ \f{(\beta + p) (2 - \beta + p) }{2 (2 + x )} t }^{-1} ,\notag\\
a (t) &\propto t^{2 (2 + x)/[(\beta + p) (2 - \beta + p) ]},
\end{align}
and hence this model describes a power-law expansion.
Note that the slow-roll parameter
\be \label{eH-ind} \epsilon_H = \f{(\beta + p) (2 - \beta + p)}{2 (2 + x)} , \ee
remains constant.
As expected, $\e_H\ll 1$ limit gives more drastic expansion since $a\propto t^{1/\e_H}$.

Since the analytic solution~\eqref{sol1-ind} for the scale factor takes a power-law form,
one may remind the power-law inflation~\cite{Abbott:1984fp,Lucchin:1984yf}, which does not satisfy the observational constraints. 
However, 
it is not necessarily the case for the solution \eqref{sol1-ind} since the consistency relations between inflationary spectral parameters and model parameters are in general modified by the existence of non-minimal coupling. 
Indeed, we shall show below that the model has observationally viable parameter region.

The normalization factor $b_i$ and sound speed $c_i$ in the tensor and scalar quadratic actions~\eqref{bc} read
\begin{align} \label{bcind}
b_t &= \xi \mk{ \e_H t }^{4/(\beta + p)},  && 
c_t^2 = 1 ,\notag\\
b_s &= \f{\xi (4 + \beta + p) }{\beta + p}\mk{ \e_H t }^{4/(\beta + p)} , &&
c_s^2 = \f{(2+x)(4 + \beta + p)}{x(2-\beta+p)},
\end{align}
among which in particular $c_t$ and $c_s$ are constant.
It is worthwhile to note a simple relation holds
\be \label{ebc} \e_H=\f{(\beta+p)^2}{2x}\f{b_s /b_t }{c_s^2/c_t^2} . \ee
Since $\epsilon_H, c_t, c_s$ are constant and $b_t, b_s$ are power law function of $t$, the slow-roll hierarchy is simple:
\begin{align} \label{srhi-ind}
\d_1 &\equiv \f{1}{2}\f{d\ln \e_H}{dN} - \e_H = - \e_H,\notag\\ 
\d_{j+1} &\equiv \f{d\d_j}{dN} + \d_j(\d_1 - j\e_H) 
=  (j+1)!\, (-\e_H)^{j+1} , \quad (j\geq 1),\notag\\
\xi_{t1} &\equiv \f{d\ln b_t}{dN} 
= \f{4\e_H}{\beta + p}, \notag\\
\xi_{s1} &\equiv \f{d\ln b_t}{dN} 
= \f{4\e_H}{\beta + p},
\end{align}
all of which are constant.
Here, $N\equiv \ln a$ is the number of e-foldings.
Note that higher order slow-roll parameters $\xi_{t,j}, \xi_{s,j}$ with $j\geq 1$ identically vanish as $\xi_{t1}, \xi_{s1}$ are constant.
Also, we can neglect $\delta_{j+1}$ with $j\geq 1$ so long as we assume $\e_H\ll 1$ and focus on the leading order $O(\e_H)$.

With \eqref{sol1-ind} and \eqref{bcind}, the time-varying modes of the superhorizon solutions \eqref{shsol} scale as 
\begin{align} 
\int dt \f{c_t^2}{a^3 b_t} \propto 
a^{p_t}  ,\quad
\int dt \f{c_s^2}{a^3 b_s \e_H} \propto a^{p_s}  ,
\end{align}
where 
\begin{align} 
p_t =p_s= 
-3-\beta -\f{2-\beta+p}{2+x}. 
\end{align}
Therefore, for the parameter values where the right-hand side is positive, these modes are growing.
On the other hand, if they are decaying mode, 
we can apply the standard picture where scalar and tensor perturbations are frozen outside the horizon.
In this case, by focusing on slow-roll regime and considering the leading order in $\e_H$, we obtain~\cite{Motohashi:2017gqb}
\begin{align}
\Delta_\zeta^2 &\approx \f{H^2}{8 \pi^2 b_s c_s \e_H} \Bigg|_{ks=1}, \notag\\
n_s-1 &\approx - 4\e_H - 2\d_1 - \xi_{s1} \big|_{ks=1} , 
\label{pss}
\end{align}
for scalars, and
\begin{align}
\Delta_{\gamma}^2 &\approx \f{H^2}{2 \pi^2 b_tc_t} \Bigg|_{k\eta=1}, \notag\\
n_t &\approx - 2\e_H - \xi_{t1} \big|_{k\eta =1},
\label{pst}
\end{align}
for tensors, where the conformal time $\eta$ and scalar sound horizon $s$ are defined by 
\begin{align}
\eta &\equiv \int^{t_{\rm end}}_t \f{dt}{a} = \f{1}{1-\e_H^{-1}} \mk{t^{1-\e_H^{-1}}_{\rm end} - t^{1-\e_H^{-1}} } \simeq \e_H t^{-1/\e_H},\notag\\
s &\equiv \int^{t_{\rm end}}_t \f{c_s dt}{a} = c_s \eta .
\end{align}

Since $\e_H, \d_1, \xi_{s1}, \xi_{t1}$ are constants from \eqref{eH-ind} and \eqref{srhi-ind}, the scalar and tensor tilts in \eqref{pss} and \eqref{pst} are also constants and simply given by
\begin{align} \label{nseq}
n_s-1 = n_t = 
2 \beta - \f{4 (2 - \beta + p)}{2 + x}.
\end{align}
Therefore, the scalar and tensor power spectra have the same tilt in this model.

On the other hand, $H, b_s, b_t$ in the scalar and tensor power spectra are time dependent, and one needs to be careful for the evaluation of the tensor-to-scalar ratio. 
Since the scalar and tensor sound speeds are different, the mappings from inflaton field value to wavenumbers for scalar and tensor power spectra are different.
Therefore, evaluating the scalar and tensor power spectra at $k\eta=1$ leads to an incorrect evaluation of tensor-to-scalar ratio in general.  
In our case we can make use of the analytic solutions and explicitly write down the power spectrum evaluated at $ks=1$ and $k\eta=1$ for scalar and tensor perturbation respectively:
\begin{align} \label{sol1-D2}
\Delta_\zeta^2
&\approx \f{\beta+p}{8 \pi^2 c_s\e_H\xi(4+\beta+p)[\e_H (\e_Hc_sk)^{\e_H}]^{2\mk{1+\f{2}{\beta+p}}}} , \notag\\
\Delta_{\gamma}^2 
&\approx \f{1}{2 \pi^2 \xi [\e_H (\e_Hk)^{\e_H}]^{2\mk{1+\f{2}{\beta+p}}}}, 
\end{align}
The tensor-to-scalar ratio evaluated at the same wavenumber is given by
\be \label{req}
r\equiv \f{4\Delta_{\gamma}^2}{\Delta_{\zeta}^2} 
= 16 c_s \e_H \mk{ 1 + \f{4}{\beta+p} } c_s^{ 2\e_H \mk{1 + \f{2}{\beta+p} } }
\ee

Let us consider the condition on the model parameters $\beta,\sigma,\xi$ for viable model from theoretical and observational point of view.
First of all, they need to satisfy the no-ghost condition~\eqref{noghost-ind} and the condition~\eqref{realp} for Case~1.
In addition, we impose the following set of conditions:
\begin{enumerate}
	\item[(i)] \label{con1} $\epsilon_H>0$.
	\item[(ii)] \label{con2} $c_s^2,c_t^2, b_t, b_s>0$.
	\item[(iii)] \label{con3} $p_s, p_t<0$. 
\end{enumerate}
With the analytic solution~\eqref{sol1-ind}, the condition~(\hyperref[con1]{i}) guarantees that the Hubble parameter $H(t)$ is positive and decreasing in time.
The condition~(\hyperref[con1]{ii}) guarantees the absence of the ghost and gradient instabilities in scalar and tensor perturbations.
The condition~(\hyperref[con1]{iii}) guarantees the absence of growing mode of tensor and scalar perturbations on superhorizon scales.

First, we show that the conditions~(\hyperref[con1]{i}) and (\hyperref[con1]{ii}) dramatically reduces the parameter space. 
From the relation~\eqref{ebc}, the conditions~(\hyperref[con1]{i}) and (\hyperref[con1]{ii}) is equivalent to require $x\equiv \sigma/\xi>0$.
Combined with the fact that $b_t>0$ means $\xi>0$, we obtain $\sigma=+1$.
For $x>0$, the no-ghost condition~\eqref{noghost-ind} and the assumption $x\ne -2$ are automatically satisfied.
Under $x>0$, the remaining conditions $b_s>0$ and $c_s^2>0$ read
\be \label{con1a2} \f{4 + \beta + p}{\beta + p} >0 ~~\text{and}~~
\f{4 + \beta + p}{2-\beta+p} > 0. \ee
For $\beta<0$, the condition~\eqref{con1a2} simplifies as $\beta+p>0$,
which is automatically satisfied since 
\be \label{eq1} \beta+p = \beta+\sqrt{\beta^2 - 2(4+x)\beta + 4} > \beta + |\beta| = 0. \ee
For $\beta>0$, the condition~\eqref{con1a2} simplifies as $2-\beta+p>0$, namely, 
\be \label{eq2} \sqrt{(\beta-2)^2 - 2(2+x)\beta} > \beta -2. \ee
Clearly, under $x>0$ the condition~\eqref{eq2} is not satisfied if $\beta\geq 2$.
On the other hand, the condition~\eqref{eq2} is satisfied for $\beta<2$ so long as the left-hand side is real, i.e.\ the condition~\eqref{realp} is satisfied.
Therefore, after imposing the conditions~(\hyperref[con1]{i}) and (\hyperref[con1]{ii}), the allowed parameter region is 
\be \label{param-i-ii} \sigma=+1 ~~\text{and}~~
\xi>0 ~~\text{and}~~
\beta < 4 + \xi^{-1} - \sqrt{ ( 4+\xi^{-1} )^2 - 4 } .
\ee
The last condition is obtained from the condition~\eqref{realp} and $\beta<2$.
The right-hand side of the last condition is monotonically increasing from $0$ for $\xi\to 0$ to an asymptotic value $4 - 2 \sqrt{3}\approx 0.536$ for $\xi\to \infty$, 
so $\beta< 0.54$ holds for any positive $\xi$.

\begin{figure}
	\centering
	\includegraphics[width=0.45\columnwidth]{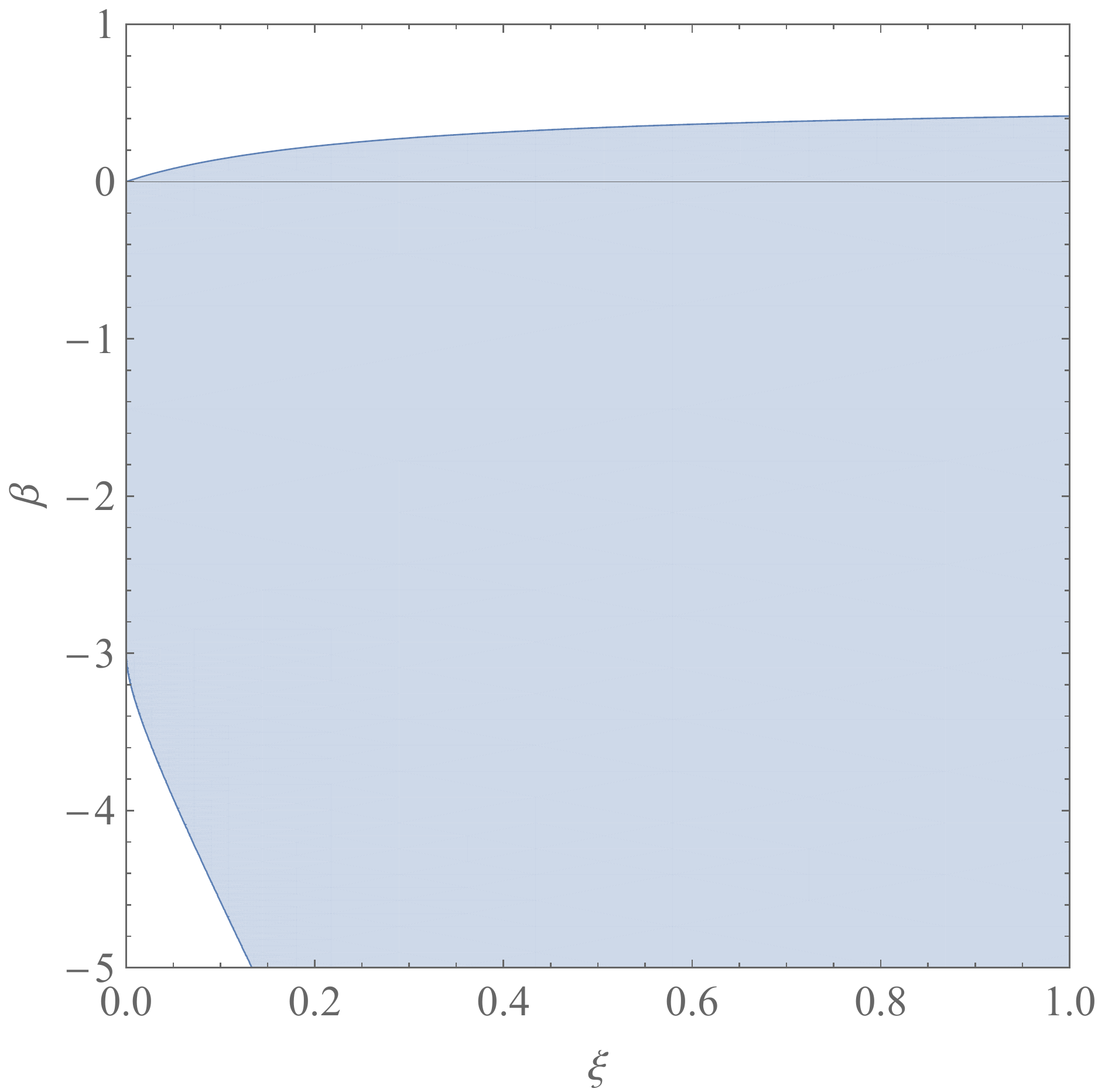}
	\caption{Parameter region~\eqref{sol1m-param} that satisfies the conditions~(\hyperref[con1]{i})--(\hyperref[con1]{iii}) for the particular solution~\eqref{h1}.}
	\label{fig:param-sol1}
\end{figure}

Second, we impose the condition~(\hyperref[con1]{iii}).
With $\xi>0$, this condition simplifies as
\be (1 + \xi^{-1}) \beta + 8 + 3 \xi^{-1} + 
\sqrt{ \beta^2 - 2 (4 + \xi^{-1}) \beta + 4 } > 0. \ee
Clearly, the condition is satisfied for $\beta>0$ so long as the left-hand side is real.
For $\beta<0$, this condition yields a lower bound.
Combining it with \eqref{param-i-ii}, we obtain 
\be \label{sol1m-param} \sigma=+1 ~~\text{and}~~
\xi>0 ~~\text{and}~~
-3 - 6 \xi (1 + \sqrt{1 + (6 \xi)^{-1}}) <\beta < 4 + \xi^{-1} - \sqrt{ ( 4+\xi^{-1} )^2 - 4 } , \ee
as the allowed parameter region where the conditions~(\hyperref[con1]{i})--(\hyperref[con1]{iii}) are satisfied,
which is depicted in Fig.~\ref{fig:param-sol1} as a shaded region.
However, one can show that there is no parameter region that satisfies the observational constraint on $(n_s,r)$ given by Planck 2018 results~\cite{Akrami:2018odb}.

\subsection{Case 1, solution 2}

Next, let us study the case of another particular solution~\eqref{h2}. 
As already mentioned above, the analytic solutions for this case are given by a replacement $p\to -p$.
Therefore, we obtain
\begin{align} \label{sol2-ind}
	V(\phi) &= \f{\xi[12 (3 - \beta - p) 
	+ \{ 20 - \beta^2 - p (4 + \beta) \} x
	+ (3 + \beta)x^2]}{(2 + x)^2 }  \phi^{2 - \beta + p}  , \notag\\
	\phi(t) &= \mk{ \e_H t }^{2/(\beta - p)} ,\notag\\
	H(t) &= \mk{ \e_H t }^{-1} ,\notag\\
	a (t) &\propto t^{\e_H^{-1}},\notag\\
	\epsilon_H &= \f{(\beta - p) (2 - \beta - p)}{2 (2 + x)} ,
\end{align}
and 
\begin{align} \label{sol2-bcind}
	b_t &= \xi \mk{ \e_H t }^{4/(\beta - p)},  && 
	c_t^2 = 1 ,\notag\\
	b_s &= \f{\xi (4 + \beta - p) }{\beta - p}\mk{ \e_H t }^{4/(\beta - p)} , &&
	c_s^2 = \f{(2+x)(4 + \beta - p)}{x(2-\beta-p)}.
\end{align}
Again, we have a simple relation 
\be \label{sol2-ebc} \e_H=\f{(\beta-p)^2}{2x}\f{b_s /b_t }{c_s^2/c_t^2} . \ee
With the slow-roll hierarchy given by \eqref{srhi-ind} with the replacement $p\to -p$, we obtain
\begin{align} 
p_s = p_t &= -3-\beta -\f{2-\beta-p}{2+x}, \notag\\
n_s-1 = n_t &= 2 \beta - \f{4 (2 - \beta - p)}{2 + x}, \notag\\
r& 
= 16 c_s \e_H \mk{ 1 + \f{4}{\beta-p} } c_s^{ 2\e_H \mk{1 + \f{2}{\beta-p} } }
\end{align}
whereas the amplitude of the spectra are given by \eqref{sol1-D2} with the replacement $p\to -p$.

\begin{figure}
	\centering
	\includegraphics[width=0.45\columnwidth]{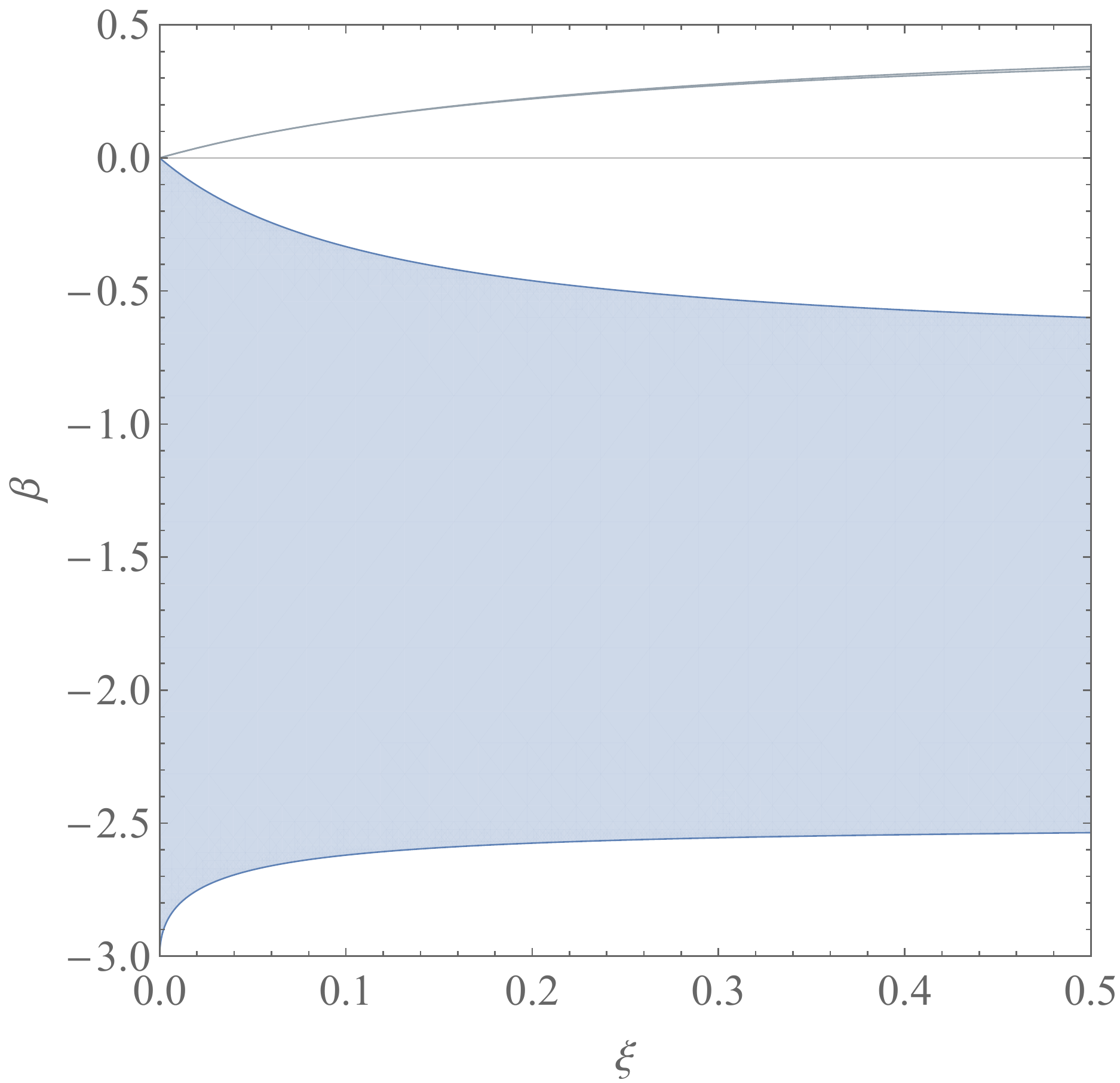}
	\includegraphics[width=0.45\columnwidth]{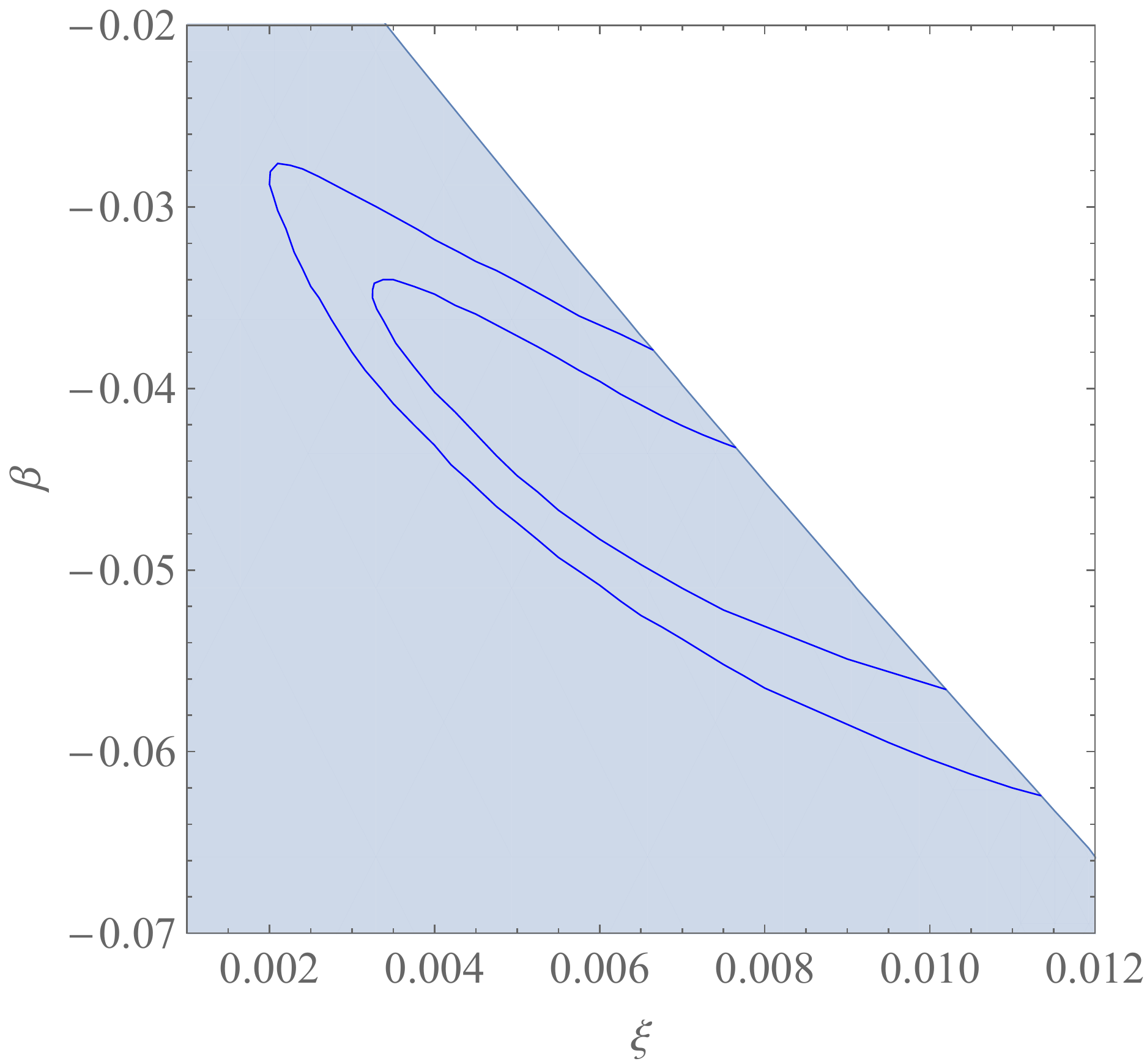}
	\caption{
	Left: Parameter region~\eqref{sol1p-param} that satisfies the conditions~(\hyperref[con1]{i})--(\hyperref[con1]{iii}) for the particular solution~\eqref{h2}.	
	Right: Observational constraint on $(n_x,r)$~\cite{Akrami:2018odb} projected on model parameter space.}
	\label{fig:param-sol2}
\end{figure}

We then impose the condition~(\hyperref[con1]{i})--(\hyperref[con1]{iii}) and observational constraint on $(n_s,r)$, and identify allowed parameter region.
Requiring the same condition~(\hyperref[con1]{i})--(\hyperref[con1]{iii}), we obtain the viable parameter region
\be \label{sol1p-param} \sigma=+1 ~~\text{and}~~
\xi>0 ~~\text{and}~~
\kk{ 
	-3 -6\xi ( 1-\sqrt{1+(6 \xi)^{-1}}  ) < \beta < -\f{6}{8+\xi^{-1}} ~~\text{or}~~
	\f{2}{4+\xi^{-1}} < \beta < 4 + \xi^{-1} - \sqrt{ ( 4+\xi^{-1} )^2 - 4 }    
}, \ee
which is depicted in the left panel of Fig.~\ref{fig:param-sol2} as a shaded region.
One can show that the potential is always positive for this parameter region.
Let us first focus on the $\beta>0$ part of the viable parameter region~\eqref{sol1p-param}.
For small $\xi$, the last condition is approximately
\be 2 \xi - 8 \xi^2 + 32 \xi^3 + O(\xi^4) < \beta < 2 \xi - 8 \xi^2 + 34 \xi^3 + O(\xi^4). \ee
Therefore it requires a fine-tuning for $\beta$.
Furthermore, one can show that for this region, the analytic solution describes the inflaton climbing up the potential as $(2 - \beta + p) (\beta - p)>0$, which is not an attractor solution.
In general, one needs a transition from the non-attractor regime to attractor regime 
by connection to a different potential 
or changing the field trajectory to a direction for a different scalar field, 
which is sufficiently massive during the non-attractor regime.
Here, we focus on the $\beta<0$ part of the parameter region~\eqref{sol1p-param}. 
In this case the inflaton rolls down the positive and increasing potential.

Finally, we compare our model with the particular solution~\eqref{h2} with the latest observational constraint on $(n_x,r)$ by Planck 2018 results~\cite{Akrami:2018odb}. 
The right panel of Fig.~\ref{fig:param-sol2} depicts the viable parameter region, where two curves amount to the 68\% and 95\% confidence regions.

\subsection{Case 2}

Next we consider the Case 2 that does not satisfy \eqref{realp},
for which 
$p$ is pure imaginary, 
and hence
we denote 
\be p= iq = i \sqrt{-[(\beta-2)^2 - 2 \beta ( 2 + \sigma/\xi)] } \ee
with $q$ being real.
In this case we take
\be \label{sol-realq} h_1(\phi) = \phi^{-\beta/2} \cos \mk{\f{q\ln\phi}{2}}, \quad 
h_2(\phi)=\phi^{-\beta/2} \sin \mk{\f{q\ln\phi}{2}}. \ee
The general solution for the Hubble parameter $H(\phi)$ is given by \eqref{Hsol} with these $h_1(\phi),h_2(\phi)$ with real integration constants $c_1,c_2$.
The relation between $\phi$ and $t$ can be obtained by solving \eqref{dphi}, 
which is again expressed in terms of ${}_2F_1$.
Then $\phi(t)$ can be obtained by solving \eqref{dphi}, and the potential is given by \eqref{Vsol}.  
In this case the particular solution does not give simple solution and hence we simply provide the potential for general solution
\begin{align}
v_1(\phi) &= \f{\xi \phi^{2 - \beta}}{2 (2 \xi + \sigma)^2} [\gamma_1 \cos(q \ln \phi) + \gamma_2 \sin(q \ln \phi) + \gamma_3 ] ,\notag\\
v_2(\phi) &= \f{\xi \phi^{2 - \beta}}{2 (2 \xi + \sigma)^2} [-\gamma_1 \cos(q \ln \phi) - \gamma_2 \sin(q \ln \phi) + \gamma_3 ] ,\notag\\
v_3(\phi) &= \f{\xi \phi^{2 - \beta}}{2 (2 \xi + \sigma)^2} [-\gamma_2 \cos(q \ln \phi) + \gamma_1 \sin(q \ln \phi) ] ,
\end{align}
where
\begin{align}
\gamma_1 &\equiv 3 + \beta +  12 (3 - \beta) \xi^2 + (20 - \beta^2) \xi \sigma , \notag\\
\gamma_2 &\equiv q \xi [12 \xi + (4 + \beta) \sigma] , \notag\\
\gamma_3 &\equiv (3 - \beta) (1 + 12 \xi^2 + 8 \xi \sigma) .
\end{align}

\begin{figure}
	\centering
	\includegraphics[width=80mm]{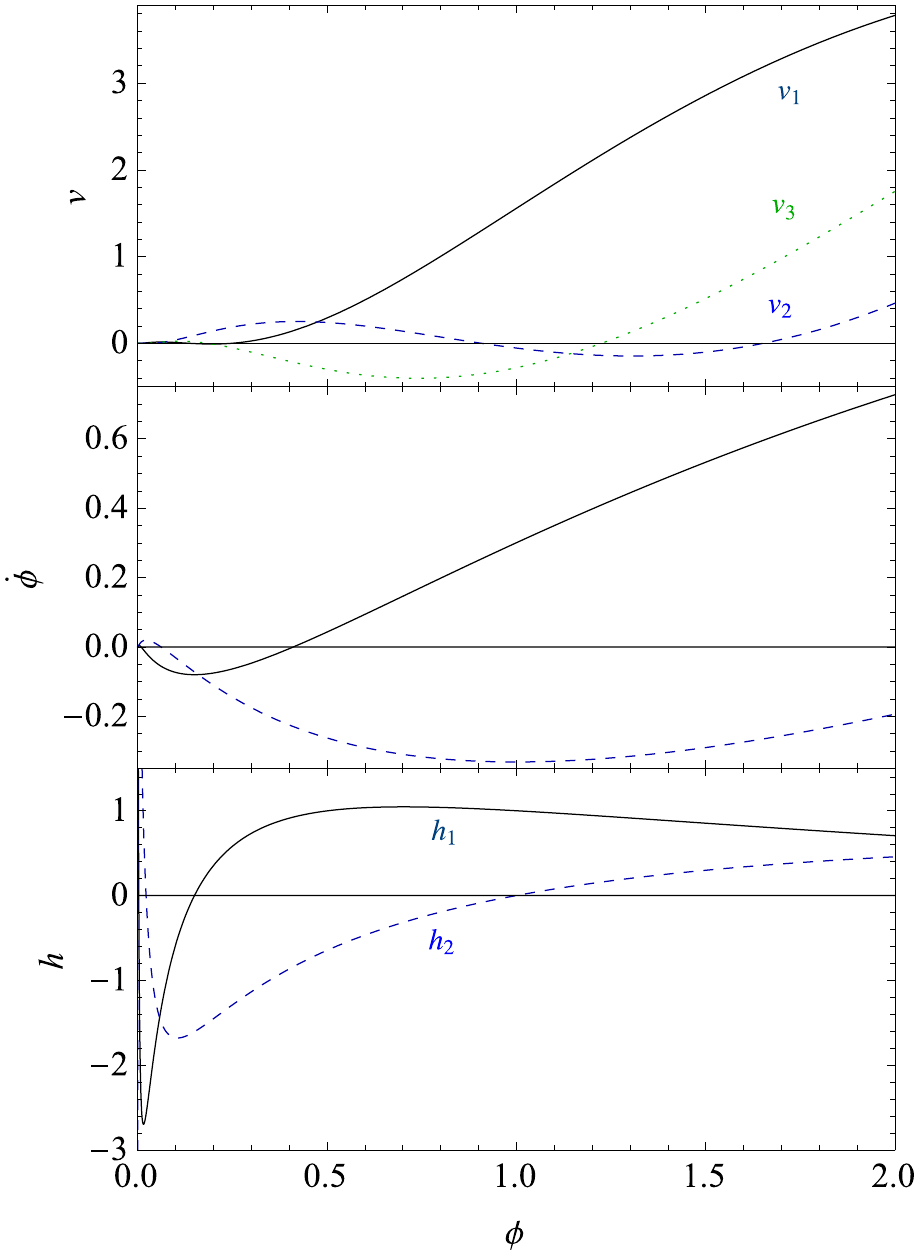}
	\caption{The Case 2. Hubble parameter $h_1(\phi)$, $h_2(\phi)$, and the potential $v_1(\phi)$,  $v_2(\phi)$, $v_3(\phi)$, and the phase space diagram for $\sigma=1, \xi=1/3, \beta = 1/2$.  }
	\label{fig:vh-ind}
\end{figure}

Exact solutions are depicted in Fig.~\ref{fig:vh-ind}. 
We are interested in a region where the Hubble constant is positive and decreasing, keeping its variation small.
Since the difference of $h_1(\phi)$ and $h_2(\phi)$ is only phase, 
let us focus on $h_1(\phi)$ in Fig.~\ref{fig:vh-ind}.
There indeed exists a region for the positive and slow variation of $h_1(\phi)$, and since $\dot\phi>0$, 
the Hubble parameter is decreasing for $\phi\gtrsim 0.5$.
However, this solution corresponds to the potential $v_1(\phi)$, and it is increasing as $\phi$ evolves,
which implies that the inflaton is climbing up the potential.
Such a case is not an attractor solution, and in general requires a transition from non-attractor regime to attractor regime.

\section{Non-minimally coupled scalar field}
\label{sec:st}

Let us consider another special case of scalar-tensor gravity -- GR with a non-minimally coupled scalar field as a source. Then 
\be f(\phi) = 1 - \xi\phi^2. \ee
In this case, slow-roll inflationary solutions was first considered in~\cite{Spokoiny:1984bd,Futamase:1987ua}.
Since $\omega_{\rm BD}(\Phi) = \sigma f/f'^2 = \sigma(1-\xi\phi^2) /4\xi^2\phi^2$ in this case, the ghost-free parameter region \eqref{nogh} reads
\be -6 < \f{\sigma(1-\xi\phi^2)}{\xi^2\phi^2} < \infty . \ee
Clearly, the inequality does not hold for $\phi=0$. 
Therefore, so long as we focus on ghost-free theory, we should cut the potential before the inflaton reaches $\phi = 0$.  For $\phi\neq 0$, allowed parameter regions are
\begin{align}
\sigma=+1&: \quad  \xi(6\xi-1) \phi^{2} > -1, \notag\\
\sigma=-1&: \quad \xi(6\xi+1) \phi^{2} > 1.
\end{align}
The inequality should be satisfied for field region of interest.
For the first case, the inequality is satisfied for any $\phi$ if $0<\xi < 1/6$.
Note also that the condition $\sigma+f''\ne 0$ reads $\xi\ne \sigma/2$.

The differential equation~\eqref{dH} reads
\begin{align} 
\label{dH-1} 
& (1-\xi\phi^2) H'' - (\beta+1) \xi \phi H' + \kk{  ( 1 - 2\beta )\xi + \f{1}{2} \sigma\beta } H = 0. 
\end{align}
Note that for the limit $\xi\to 0$ we recover the canonical constant roll model in~\cite{Motohashi:2014ppa}, and two independent solutions are given by $H(\phi)\sim e^{\pm \sqrt{\f{\beta}{2}}\phi}$.
With $\xi\neq 0$ two independent solutions for this equation are given by the associated Legendre function $P_n^m(x)$ and $Q_n^m(x)$,
\be \label{solpq} 
H(\phi) = 
c_1 ( \xi \phi^2 - 1 )^{-m/2} P_n^m (\sqrt{\xi} \phi) 
+ c_2 ( \xi \phi^2 - 1 )^{-m/2} Q_n^m (\sqrt{\xi} \phi), \ee
where
\be 
n\equiv -\f{1}{2} + \f{1}{2} \sqrt{4 + \beta^2 + 2 \beta \mk{ \f{\sigma}{\xi} -4 }} , \quad 
m\equiv \f{1}{2} (\beta-1) . \ee

Plugging the solution~\eqref{solpq} for $H(\phi)$ into \eqref{dphi}, we can write down $\dot\phi$ as a function of $\phi$, with which 
we can write down the potential \eqref{pot}.

\subsection{Special case}

Let us consider a simple case with $\beta=1$, for which we have $m=0$ and $n = (-1 + \sqrt{2\sigma/\xi-3}) /2$.
Furthermore, if $n$ is a natural number, the solution simplifies dramatically.

The case $n=0$ corresponds to $\sigma/\xi=2$, i.e.\ $\sigma+f''=0$, so instead of the solution \eqref{solpq} we use \eqref{sol2} and hence only de Sitter solution is allowed for $\beta=1$, which is the solution \eqref{dSb1}.

The case $n=1$ corresponds to $\sigma/\xi=6$, for which \eqref{solpq} reads
\be 
H(\phi) = c_1 \sqrt{\f{\sigma}{6}}\phi
+ c_2 \kk{ -1 + \sqrt{\f{\sigma}{6}} \arctan \mk{\sqrt{\f{\sigma}{6}} \phi}   }.
\ee

Let us focus on the particular solution with $c_2=0$.
The Hubble parameter positive for $c_1\sqrt{\sigma} >0$.
In this case the 
The potential~\eqref{pot} is then given by 
\be V(\phi) = -\f{1}{16} c_1^2 (3 \phi^4- 20 \sigma \phi^2 + 12) . \ee
By solving \eqref{conroll}, the analytic solution for the inflaton is given 
\be  \phi(t) = \f{2}{c_1} \sqrt{\f{6}{\sigma}} \tan t.  \ee
Plugging it into $H = \dot a/a = c_1 \sqrt{\sigma/6}\, \phi$, we obtain 
\be H(t)=2\tan t, \quad a(t) \propto \f{1}{\cos^2 t} .  \ee

\section{Exponential coupling}
\label{sec:exp}

One more case when the equation~\eqref{dH} can be solved in terms of standard higher transcendental function is 
\be f(\phi) = \exp(C\phi). \ee 
In this case, \eqref{dH} reads
\be 2 e^{C \phi} (C^2 e^{C \phi} + \sigma) H'' 
+ C e^{C \phi} [C^2 e^{C \phi} (-1 + \beta) + (1 + \beta) \sigma] H' 
+ [C^4 e^{C \phi} \beta + C^2 e^{C \phi} (-1 + 3 \beta) \sigma + \beta \sigma^2] H = 0. \ee

For the case with $\sigma=0$, the solution is given by
\be H(\phi) = c_1 \exp\kk{ \f{C}{4} (1 - \beta + \sqrt{1 - 10 \beta + \beta^2} ) \phi } 
+ c_2 \exp\kk{ \f{C}{4} (1 - \beta - \sqrt{1 - 10 \beta + \beta^2} ) \phi } 
\ee

To derive analytic solution for general case with $\sigma\ne 0$, it is convenient to introduce a function $Z(\phi)\equiv\int H d\phi=\dot\phi/\beta$.
First, we integrate \eqref{dH} to obtain a second-order equation with respect to $Z(\phi)$.
Then, regarding $Z$ as a function of $x=\exp(C\phi)$, we obtain 
\be \f{d^2Z}{dx^2} + \f{1 + \beta}{2 x} \f{dZ}{dx} + \f{\beta (x + \sigma C^{-2})}{2 x^3} Z = 0 . \ee
Further, defining $y=\sqrt{\f{2\sigma\beta}{C^2 x}}$, which is real or pure imaginary depending on $\sigma\beta \gtrless 0$, we arrive at
\be \f{d^2Z}{dy^2} 
+ \f{2 - \beta}{y} \f{dZ}{dy} + \mk{ 1 + \f{2 \beta}{y^2} } Z = 0. \ee
The solution can be written as
\be Z = c_1 y^{(\beta-1)/2} J_\nu(y) + c_2 y^{(\beta-1)/2} Y_{\nu}(y), \ee
where $\nu= \f{1}{2}\sqrt{\beta^2 - 10 \beta + 1}$
and $J_\nu(y), Y_\nu(y)$ are the Bessel functions of the first and second kind, respectively.
The Hubble parameter $H(\phi)$ can be obtained by $H=dZ/d\phi$.
The time evolution of the scalar field $\phi(t)$ is governed by $\dot\phi = \beta Z$, with which the potential $V(\phi)$ can be obtained by \eqref{pot}.

\subsection{Special case}

As a simple case, let us consider $\beta=10$ for which $\nu=1/2$ and the Bessel functions can be written down in terms of trigonometric functions.
The Hubble parameter is then given by
\begin{align} 
H(\phi) = 
80 C^{-3} \sqrt{\f{10}{\pi}} e^{-5 C \phi/2}  
\left[ - c_1 \ck{ 5 \f{\sqrt{\sigma}}{C} \cos\mk{ 2 \sqrt{5} e^{-C \phi/2} \f{\sqrt{\sigma}}{C} } 
+ 2 \sqrt{5} e^{C \phi/2} \sin\mk{ 2 \sqrt{5} e^{-C \phi/2} \f{\sqrt{\sigma}}{C} } } \right.\notag\\
+ \left. c_2 \ck{ 2 \sqrt{5} e^{C \phi/2} \cos\mk{ 2 \sqrt{5} e^{-C \phi/2} \f{\sqrt{\sigma}}{C} } 
- 5 \f{\sqrt{\sigma}}{C} \sin\mk{ 2 \sqrt{5} e^{-C \phi/2} \f{\sqrt{\sigma}}{C} } }  \right]  .
\end{align}
For $\sigma=-1$ case, it is written in terms of hyperbolic functions and simplifies as
\be H(\phi) = 400 \sqrt{\f{2}{\pi}} e^{-5 \phi/2} \kk{  \tilde c_1 (2 e^{\phi/2} + \sqrt{5} ) e^{2 \sqrt{5} e^{-\phi/2} }  
+ \tilde c_2 (2 e^{\phi/2} - \sqrt{5} ) e^{-2 \sqrt{5} e^{-\phi/2} } }, \ee 
where we set the normalization of the scalar field as $C=1$, and redefine integration constants as $\tilde c_1 = (-i c_1 + c_2)/2$ and $\tilde c_2 = (i c_1 + c_2)/2$.
By taking $\tilde c_1$ and $\tilde c_2$ real, we obtain real $H(\phi)$. 
The potential is then given by
\begin{align} 
V(\phi) &= \f{320000}{\pi} e^{-4 \phi} 
[ - \tilde c_1^2 e^{4 \sqrt{5} e^{-\phi/2}} (48 e^\phi + 18 \sqrt{5} e^{\phi/2} - 65) \notag\\
&~~~ - 2 \tilde c_1\tilde c_2 (48 e^\phi - 35) 
+ \tilde c_2^2 e^{-4 \sqrt{5} e^{-\phi/2}} (-48 e^\phi + 18 \sqrt{5} e^{\phi/2} + 65) ] .
\end{align}

\section{Conclusion}
\label{sec:con}

We have generalized the notion of constant-roll inflation earlier introduced in GR and $f(R)$ gravity to inflationary models in more general scalar-tensor theory in which a scalar field is non-minimally coupled to gravity.
We have found in \S\ref{sec:ds} that the constant-roll condition allows a novel exact de Sitter solution with a scalar field varying with time as an analytic solution of the equations of motion.
We have found the general form of the non-minimal coupling function $f(\phi)$ for which such solution can exist.
In \S\ref{sec:ig}, by considering the constant-roll inflation in the induced gravity, we have found a simple analytic solution with the power-law inflation, 
for which we have identified a viable parameter region compatible with the latest observational constraint on $(n_s,r)$.  
In \S\ref{sec:st} and \S\ref{sec:exp}, we have also provided analytic solutions for the constant-roll inflation in the other specific cases of scalar-tensor gravity: GR driven by a non-minimally coupled scalar field.
We do not consider the stability of all specific exact solutions found in the paper 
since this requires abandoning the constant-roll assumption and, thus, requires separate consideration.

For the constant-roll inflation in the induced gravity, the observational constraint on $(n_s,r)$ restricts the constant-roll parameter as $-0.06\lesssim \beta\lesssim -0.03$.
The requirement of small $|\beta|$ is consistent with the results obtained in the previous constant-roll models~\cite{Motohashi:2014ppa,Motohashi:2017aob,Motohashi:2017vdc}.
Since we started from general constant-roll condition and found that small $|\beta|$ is needed to satisfy observational constraints, these results imply that the slow-roll condition is not a theoretical assumption taken for simplicity but actually the requirement from observation.

The constant-roll de Sitter solution found in the present paper can be exploited for inflation by assuming an appropriate mechanism for generation of primordial perturbations and transition to reheating regime.
Unlike other constant-roll scenarios, in this case $|\beta|$ is allowed to take a large amplitude, and hence the de Sitter expansion can be driven by non-slow-roll constant-roll inflaton.
It would be interesting to investigate this scenario and its implications more in detail that we leave for a future work.

\begin{acknowledgments}
H.M.\ was supported in part by 
Japan Society for the Promotion of Science (JSPS) Grants-in-Aid for Scientific Research (KAKENHI) No.\ JP17H06359 and No.\ JP18K13565.
A.S.\ was partially supported by the grant RFBR 17-02-01008. 
\end{acknowledgments}

\bibliography{ref-STconroll}

\end{document}